\documentclass[amsmath,amssymb,reprint,twocolumn,superscriptaddress]{revtex4-1}

\usepackage[utf8]{inputenc}
\usepackage[T1]{fontenc}
\usepackage[english]{babel}

\usepackage{amsmath,amssymb,amsfonts}
\usepackage{amstext, mathrsfs, textcomp}
\usepackage{mathptmx}
\usepackage{bigints}  
\usepackage{nicefrac}
\usepackage{multirow}
\usepackage{dcolumn}
\usepackage{bm}

\usepackage{changes}   
\usepackage[colorlinks=true, citecolor=blue, urlcolor=blue, linkcolor=blue]{hyperref} 

\usepackage{subfigure}
\usepackage{graphicx}
\usepackage{xcolor}
\usepackage[export]{adjustbox}

\graphicspath{{figures/}}

\begin{document}
\title{Monitoring MBE substrate deoxidation via RHEED image-sequence analysis by deep learning}

\author{\firstname{Abdourahman} \surname{Khaireh-Walieh}}
\affiliation{LAAS-CNRS, Université de Toulouse, CNRS, UPS, F-31400 Toulouse, France}

\author{\firstname{Alexandre} \surname{Arnoult}}
\affiliation{LAAS-CNRS, Université de Toulouse, CNRS, UPS, F-31400 Toulouse, France}

\author{\firstname{Sébastien} \surname{Plissard}}
\affiliation{LAAS-CNRS, Université de Toulouse, CNRS, UPS, F-31400 Toulouse, France}

\author{\firstname{Peter R.} \surname{Wiecha}}
\email[e-mail~: ]{pwiecha@laas.fr}
\affiliation{LAAS-CNRS, Université de Toulouse, CNRS, UPS, F-31400 Toulouse, France}


\begin{abstract}
	Reflection high-energy electron diffraction (RHEED) is a powerful tool in molecular beam epitaxy (MBE), but RHEED images are often difficult to interpret, requiring experienced operators.
	We present an approach for automated surveillance of GaAs substrate deoxidation in MBE reactors using deep learning based RHEED image-sequence classification.
	Our approach consists of an non-supervised auto-encoder (AE) for feature extraction, combined with a supervised convolutional classifier network.
	We demonstrate that our lightweight network model can accurately identify the exact deoxidation moment.
	Furthermore we show that the approach is very robust and allows accurate deoxidation detection during months without requiring re-training.
	The main advantage of the approach is that it can be applied to raw RHEED images without requiring further information such as the rotation angle, temperature, etc.
\end{abstract}

\maketitle
\section{Introduction}

Reflection high-energy electron diffraction (RHEED) is a widely used in-situ control method in molecular beam epitaxy (MBE) \cite{inoNewTechniquesReflection1977, horioNewTypeRHEED1996, braunAppliedRHEEDReflection1999, ichimiyaReflectionHighenergyElectron2004}.
RHEED diffraction patterns provide information about the crystal surface with atomic resolution and as the ultra high vacuum in typical growth chambers allows an easy integration of electron beam systems in MBEs, RHEED has become a standard in-situ characterization instrument in MBE, enabling unprecedented accuracy in monitoring the crystal growth. 
RHEED is highly sensitive to several key MBE parameters like the growth rate, the crystal structure, the lattice parameter and strain, etc \cite{surfacesSurfaceInterfaceCharacterization1988, joRealTimeCharacterizationUsing2018, thelanderElectricalPropertiesInAs12012, daudinHowGrowCubic1998, ohtakeStrainRelaxationInAs2020}.
However, RHEED images can be difficult to interpret since the diffraction patterns produce information in the Fourier-space. Furthermore, the actual recorded patterns are very sensitive to calibration, and often also dynamic variations in the patterns over several time-scales contain valuable information, rendering their analysis even more challenging. 
Real-time exploitation of RHEED data is therefore often limited to easily accessible information like the deposition rate. Sophisticated analysis is usually done a posteriori, on recorded RHEED images or videos.
Due to the complexity of the task, RHEED interpretation usually requires experienced operators, possessing years of machine-specific training.

A common application of RHEED is the monitoring of the native oxide removal from commercial substrates prior crystal growth.
Surface oxidation of a few nanometers due to exposure to oxygen is unavoidable during transport of epitaxial substrates, which renders their surface non-crystalline. This oxide needs to be removed before any epitaxial material deposition, which is usually done by heating. 
In the case of gallium arsenide (GaAs), the substrate is slowly heated to around 610$^{\circ}$C, while stabilizing the crystal with a constant arsenic flux of around $1.2\times 10^{-5}$\,Torr, to avoid As evaporation \cite{bastimanGaAs001Planarization2010}. 
Once the oxide is removed, in order to avoid damaging of the crystal, further temperature ramping needs to be stopped, usually temperature is in fact decreased. 
To detect the deoxidation, the MBE operator supervises the RHEED image during temperature increase, and once the diffraction pattern of a crystalline surface starts to form, the operator manually ends the heating procedure.
Not only is the constant presence of the operator required, due to its manual character the deoxidation procedure is furthermore error-prone. Automatic detection of the deoxidation is challenging, first because RHEED patterns are often weak since the raw substrate surfaces are not atomically flat, and second because the RHEED image contrast is dependent on some parameters like filament current or electron beam angle, and hence is not exactly constant in each run. Finally, the substrate is usually laying on a rotating sample holder, hence the RHEED pattern constantly changes.

\begin{figure*}[t]
	\centering
	\includegraphics[width = \textwidth]{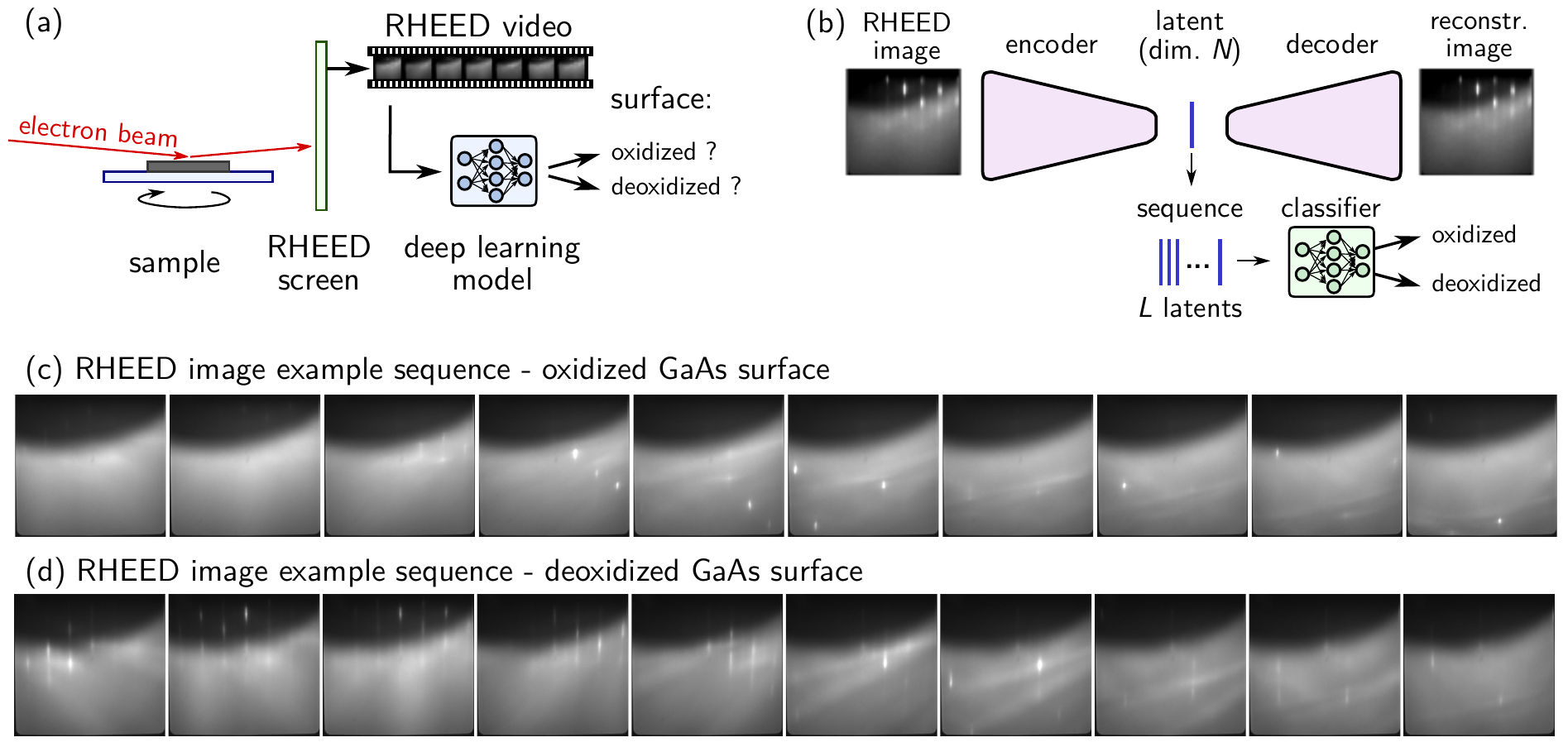}
	\caption{\textbf{Deoxidation detection problem.} 
		(a) Short sequences of the RHEED video obtained from a rotating GaAs substrate are analyzed by a deep learning neural network model (NN), to determine if the substrate deoxidation process has terminated.
		(b) The DL model consists of two stages. An autoencoder encodes every single RHEED image into a latent vector of dimension $N$ with the goal of reconstructing the original image. The latent thus contains all relevant features of the RHEED image. Sequences of $L$ latent vectors are then used for classification of deoxidized surfaces. The classifier network thus analyses short RHEED videos covering a certain rotation angle.
		(c) Example sequence of 10 consecutive raw RHEED images obtained from an oxidized GaAs surface.
		(d) Example sequence of 10 consecutive raw RHEED images obtained after deoxidation from the same GaAs surface as shown in (c).
	}
	\label{fig:setup_scheme}
\end{figure*}

Methods from artificial intelligence including deep learning are increasingly applied to nano- and material-science \cite{sachaArtificialIntelligenceNanotechnology2013, renModelingApplicationCzochralski2021, schimmelArtificialIntelligenceCrystal2022, choudharyRecentAdvancesApplications2022, battieRapidEllipsometricDetermination2022}. 
Recently, first attempts have been reported to use statistical methods and machine learning for RHEED image interpretation \cite{brownModelingMBERHEED1997, vasudevanBigDataReflectionHigh2014, provenceMachineLearningAnalysis2020, kwoenClassificationReflectionHighEnergy2020, kwoenMulticlassClassificationReflection2022}.
Inspired by these pioneering works, we propose to use a deep learning (DL) approach for classification of oxidized and deoxidized substrates via their RHEED patterns, to resolve the above described problems. 
As mentioned above, due to the sample rotation, the RHEED signal can confidently indicate deoxidation only during short moments, when the electron beam is aligned with a lattice direction of the crystal. 
In contrast to recent propositions to use DL with RHEED for surface reconstruction identification \cite{kwoenClassificationReflectionHighEnergy2020, kwoenMulticlassClassificationReflection2022}, we therefore propose a model which analyzes \textit{sequences} of RHEED patterns (i.e. videos), instead of single images. 
To this end, we propose a two-stage deep learning model: The first stage is an autoencoder, which compresses each full-resolution RHEED image into a low-dimensional latent vector. The second stage subsequently determines the oxidation state for a sequence of such latent vectors, hence for the compressed representation of a short RHEED video sequence.
We provide a detailed analysis of the required latent and sequence lengths and demonstrate the accuracy of the model as well as its stability over a period of more than 6 months between training data sampling and testing.
Finally, we provide online the data, codes as well as pre-trained models to reproduce our results \cite{khaireh-waliehRHEEDDeoxidationDataset2022}.

\section{Results and discussion}

\begin{figure*}[t]
	\centering
	\includegraphics[width = 18cm]{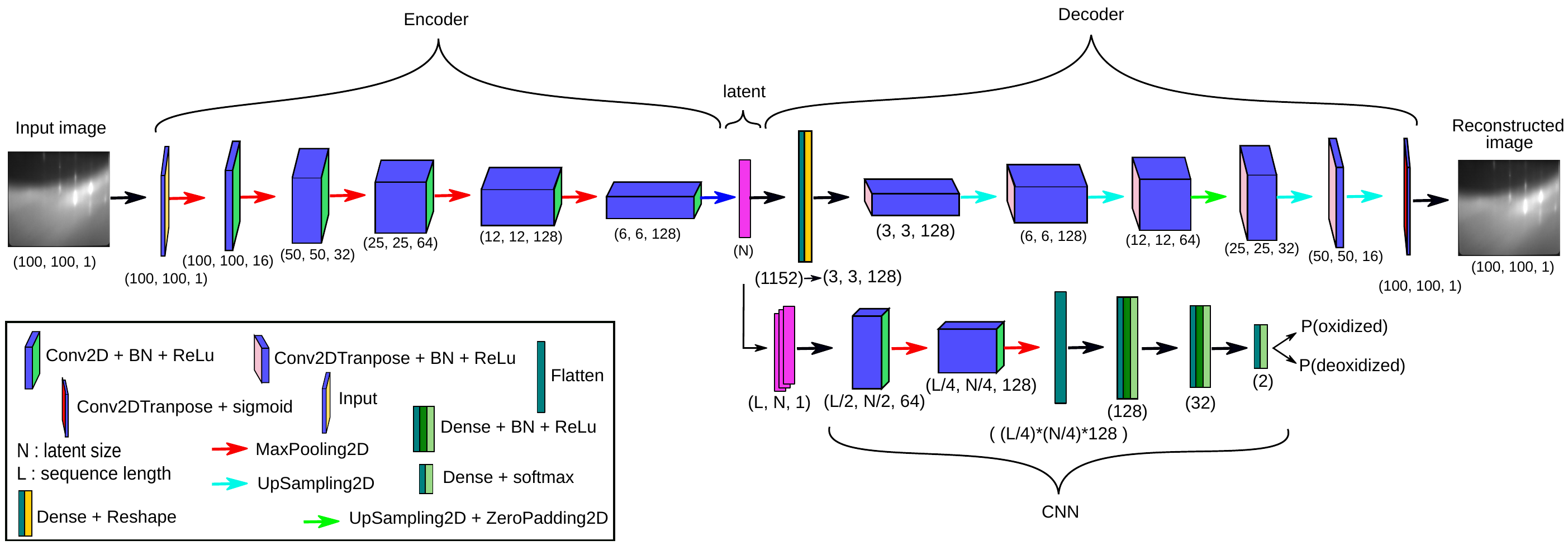}
    \caption{\textbf{Detailed network architecture.} In a CNN autoencoder (AE, top), each RHEED image is first compressed through an convolutional encoder into a 1D latent vector of length $N$. 
    Training target of the AE is reconstruction of the original image from the latent vector (through the decoder stage, only used during training).
    The second stage of the model is a classifier CNN network (bottom right), taking as input a sequence of $L$ latent vectors, corresponding to a series of RHEED images. 
    These $L$ latent vectors are stacked and passed into a CNN for classification into two classes: oxidized and deoxidized. All convolutions are followed by batch normalization (BN) \cite{ioffeBatchNormalizationAccelerating2015} and ReLU activation. For descriptions of the different network layers we refer the interested reader to relevant literature \cite{goodfellowDeepLearning2016}.}
    \label{fig:global_architecture}
\end{figure*}

\subsection{Substrate deoxidation classification Problem}

On commercial substrates, a native oxide layer of a few nanometers encapsulates the crystal surface. The RHEED electron beam does not penetrate through this oxide, and hence does not reach the crystalline lattice (of in our case GaAs). 
The electrons are thus not diffracted, but scattered. Independent of the sample rotation angle, the RHEED signal on the fluorescent screen is diffuse and no diffraction pattern occurs (c.f. Fig.~\ref{fig:setup_scheme}c).
Without oxide layer on the other hand, the RHEED electrons are diffracted by the atomic lattice of the now crystalline surface. 
However, the transition from oxidized to deoxidized is not instantaneous and during the deoxidation the classification is often difficult. Furthermore, due to the rotation of the sample, the diffraction pattern continuously changes and, especially during the deoxidation process, the pattern arises not similarly clearly for different rotation angles. Our operator classifies the surface as deoxidized when a clear diffraction pattern occurs repeatedly during at least one full rotation cycle of the substrate.

The general problem is schematically depicted in Figure~\ref{fig:setup_scheme}a. 
Our goal is to precisely determine the moment of full oxyde removal from a GaAs substrate by monitoring the RHEED pattern during the deoxidation process.

However, as mentioned above, the image dynamics due to the constant rotation of the sample is a challenge for an algorithmic evaluation. Furthermore, disordered bright spots can occur also from oxidized surfaces (see Fig.~\ref{fig:setup_scheme}c). Therefore, an algorithmic classification is not entirely trivial.
By feeding short video sequences of several consecutive RHEED images to a classification neural network, we aim at determining the oxidation state of the substrate surface, in order to reduce the necessity of human supervision of the substrate cleaning process.

\subsection{Dataset}

To train a neural network on deoxidation reconnaissance, we generate a training dataset by capturing RHEED videos before and after the oxide removal procedure.
The images are collected in real time at 24 frames per second, while the sample rotates with 12 rounds per minute, hence we capture 120 images per full rotation.
The RHEED video is thereby captured image by image, using a CMOS Camera (Allied Vision Manta G319B) with $4\times 4$ pixel binning, resulting in raw images of $416\times 444$ pixels at 12 bit grayscale intensity resolution. 
Those images are simply converted to 8 bit format and scaled to $100 \times 100$ pixels.
In total we collected videos containing a total of 7644 RHEED images from five substrate oxide removal procedures within a period of a few days. 3110 of these images correspond to deoxidized surfaces, the rest are images from GaAs surfaces which were covered by a native oxide layer.
GaAs surface oxides decompose at temperatures around $580-630^{\circ}\,C $ \cite{springthorpeMeasurementGaAsSurface1987}. On our commercial substrates we typically observe deoxidation around $610-630^{\circ}\,C $.
Substrate temperatures at which deoxidized videos where taken are slightly lower, around $550^{\circ}\,C - 600^{\circ}\,C$ (videos where recorded during ramping of the temperature). Deoxidized images where taken directly after oxide removal, at around $610^{\circ}\,C$.
During and after deoxidation, the As$_4$ pressure for surface stabilization is held at $ 1.2*10^{-5} \hspace{0.2 cm} \text{Torr}$.

We use 20\% of the dataset for validation and the remaining 80\% for training.
In addition to the oxidized and deoxidized image sets, we also captured images during the full  deoxidation procedure. These are not used during training and serve for testing of the algorithm.
RHEED images during a further deoxidation were captured around 6 months after generation of the initial dataset. These serve for an assessment of the long-term stability of the classification.

\subsection{RHEED sequence classifier network model}

Our deoxidation monitor deep learning model is composed of two stages, as depicted schematically in figure~\ref{fig:setup_scheme}b. 
A first stage is a feature extractor network, compressing the large RHEED images into compact latent vectors. This is done separately image by image. 
The second stage is the actual classification network. Its input are sequences of latent vectors, corresponding to short RHEED videos.
We implemented the models in python using keras with tensorflow as backend \cite{cholletKeras2015, tensorflow2015-whitepaper}. For pre-processing and data management we furthermore use the packages openCV, scikit-image, scikit-learn and hdf5/h5py \cite{bradskiOpenCVLibrary2000, waltScikitimageImageProcessing2014, pedregosaScikitlearnMachineLearning2011, collettePythonHDF52013}.

\subsubsection{Image feature extraction}

As feature extractor we use a deep convolutional autoencoder neural network (AE), which has been reported to offer slightly superior compression quality compared to other dimensionality reduction methods such as principal component analysis (PCA), especially at high compression rates. 
We note however, that AEs require in general more computational ressources. Thus, if computation speed is crucial, PCA could be used instead, for with only a moderate reduction in encoding performance is expected \cite{fournierEmpiricalComparisonAutoencoders2019}.

The model details of our AE are shown in the top row of figure~\ref{fig:global_architecture}. A RHEED image goes through the encoder stage, being compressed into a latent vector of dimensionality $N$.
For training, the latent vector is fed into a decoder stage, which is an exact mirror of the encoder, except for replacing convolution layers by transpose convolutions and applying zero padding if required, to maintain correct image dimension. 
Through non-supervised training, the autoencoder learns to reconstruct the unlabeled input images from their learned latent vector representation.

Please note that we optimized the network for low parameter number, in order to have a computationally efficient model. To this end we do not double the number of channels once a depth of 128 filters is reached. We also compared the architecture with a ResNet \cite{heDeepResidualLearning2015, szegedyInceptionv4InceptionResNetImpact2016}, replacing the single convolutions by residual convolutional blocks each of which employing a sequence of 3 convolutions. The performance is similar and offers no advantage in deoxidation classification. Please note that this applies to the specific here discussed problem, for other problems the slightly improved accuracy offered by a ResNet may very well be beneficial.

\subsubsection{Sequence classification CNN}

\begin{figure}[t]
	\centering
	\includegraphics[width=.9\columnwidth]{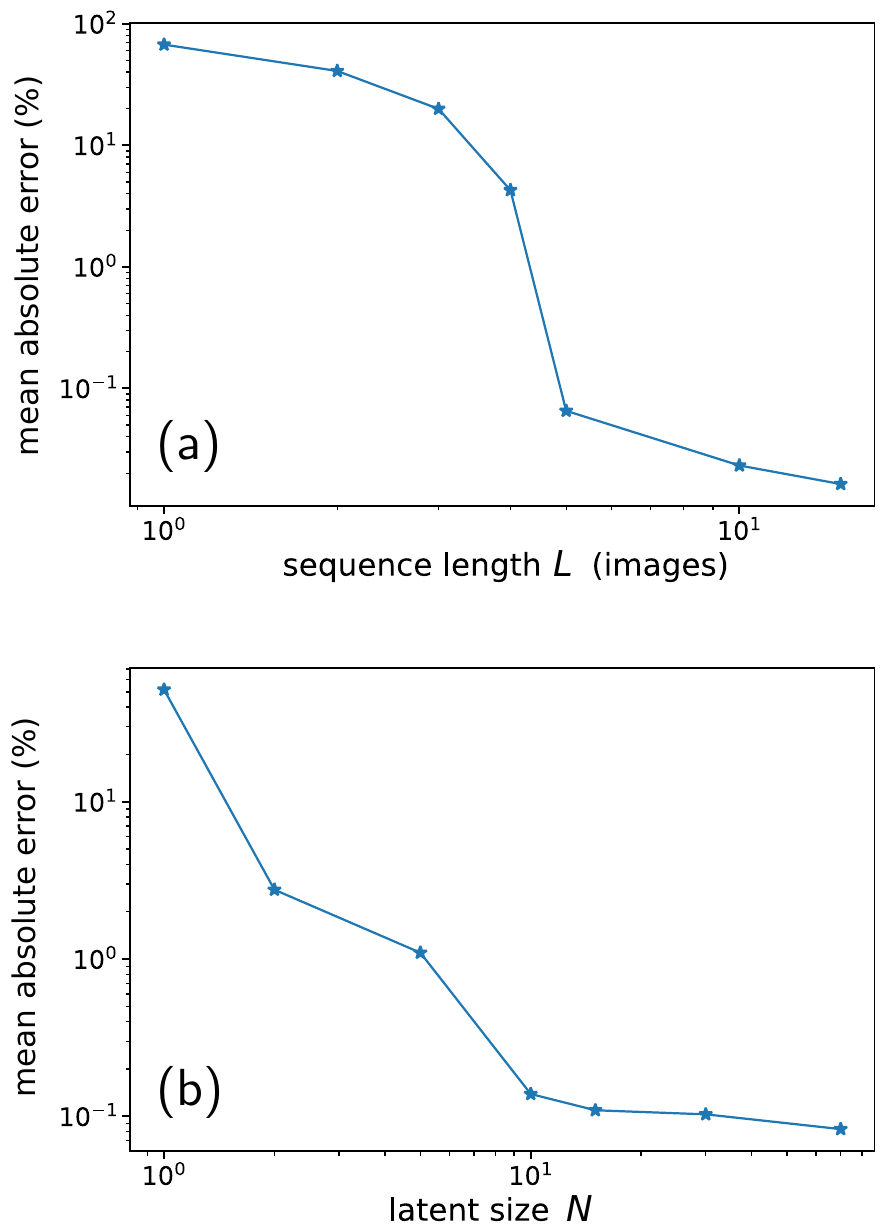}
	\caption{
	\textbf{Prediction fidelity vs. sequence length and latent size.} Mean absolute error (MAE) on the test-set for network models working with increasing sequence length and latent dimension. 
	(a) MAE as function of sequence length $L$. The latent size is fixed to $N=50$. The actual classification error of sequences $L\geq 5$ is zero.
	(b) MAE as function of latent space dimensionality $N$. The used sequence length is $L=15$. The actual classification error with latent size $N\geq 10$ is virtually zero.
	}
	\label{fig:error_vs_L_N}
\end{figure}

\begin{figure*}[t]
	\centering
	\includegraphics[width=\linewidth]{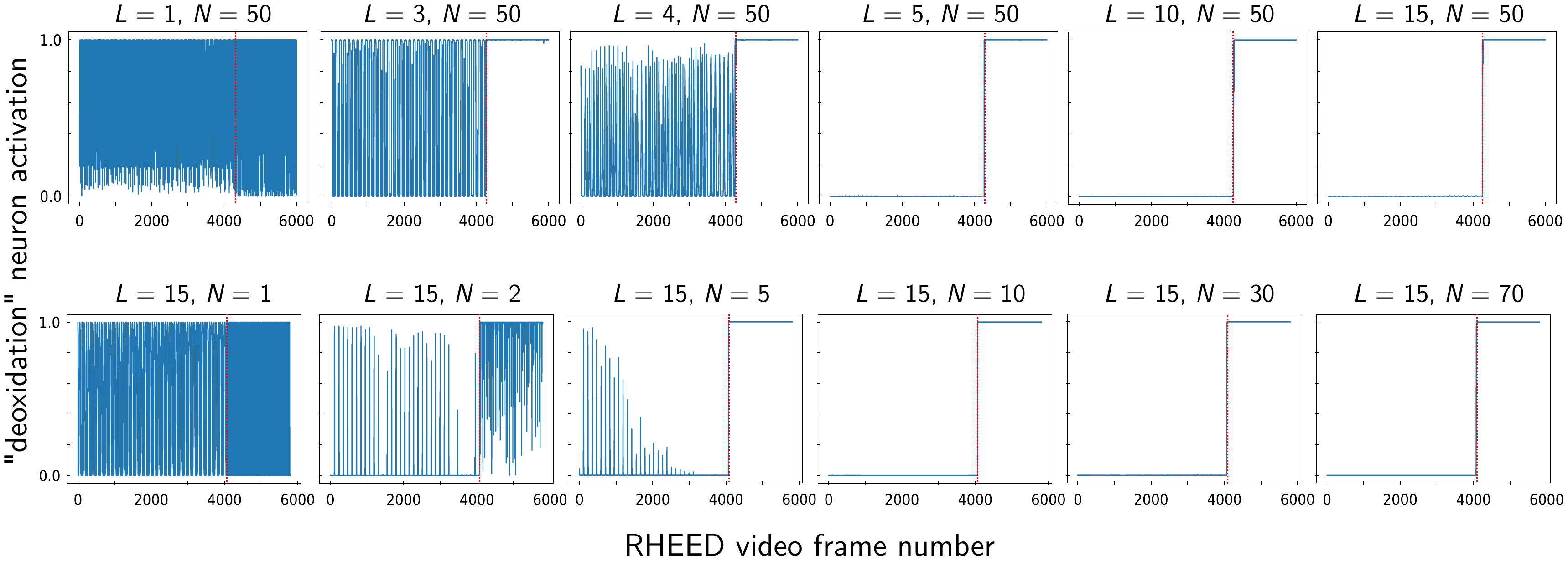}
	\caption{\textbf{Detection accuracy of deoxidation moment}. The impact of the sequence length as well as of latent vector dimension is tested on a video captured during a full deoxidation. The video consists of 31,819 RHEED images, of which the last 6,000 are shown. Deoxidation occurs around 1,800 frames before the sequence end (indicated by a red dotted line). The RHEED video is captured with 24 frames per second. Top row: Increasing sequence length $L$, the latent dimension of the autoencoder is fixed to $N=50$. Bottom row: Increasing latent dimension $N$, the sequence length of the classifier is fixed to $L=15$.}
	\label{fig:test_transition}
\end{figure*}

The second stage of our model is the actual classification network. Because the MBE sample is rotating, the RHEED images are constantly varying. Especially during deoxidation, the surface is not atomically flat and signatures of oxide removal often occur in the RHEED images only when the electron beam is aligned with the crystal lattice of the substrate.
For a high accuracy we therefore classify \textit{sequences} of RHEED images, i.e. short videos. 
To do so, we compress the RHEED video image by image using the trained AE. Using the results we then create sequences of $L$ consecutive latent vectors. These sequences are arranged as 2D arrays of size $(L,N)$, and are fed into a 2D convolutional classifier network with two output classes, one for oxidized and the second for deoxidized state of the surface.
The details of the network are shown in the bottom right of figure~\ref{fig:global_architecture}. It is trained in a supervised manner of the dataset, with the goal to predict the correct surface state from a sequence of compressed RHEED images

\subsection{Results}

\subsubsection{Autoencoder reconstruction quality}


For the tuning of the autoencoder architecture we qualitatively tested AE layouts with varying number of layers and convolutional kernels by compressing and reconstructing random RHEED images that were not used for training.
Once a layout was found that accurately reproduced the visual appearance of our RHEED images, we tested the reconstruction quality via peak signal to noise ratio (PSNR), where the original image is used as the signal and the difference between original and reconstructed image as noise. We obtained the best PSNR for $N=70$ with a value of around $90$, no further improvement was observed for larger latent dimensions. 

\subsubsection{Classification accuracy}

We test the classification accuracy of the full two-stage model (AE + classifier) with different values for the sequence length and fixed latent size $N=50$, as well as for varying latent dimensions while fixing the sequence length $L=15$.
We find that the sequence length $L$ is indeed a crucial parameter. 
Single-image classification basically fails. 
Sequences of at least $L=5$ images are required, to drop the error rates well below 1\%, which can be seen in figure~\ref{fig:error_vs_L_N}a. 
Concerning the latent size, we require at least a compression dimensionality of $N=10$ in order to get error rates well below 1\%, increasing latent dimension further improves the accuracy only marginally. This is shown in figure~\ref{fig:error_vs_L_N}b.

Subsequently we test whether the network is capable to determine the exact moment of deoxidation on a set of images captured during the entire deoxidation procedure. 
First we fix the latent size to $N=50$ and increase the sequence length successively from $L=1$ to $L=15$ (figure~\ref{fig:test_transition} top row). In agreement with the former test (figure~\ref{fig:error_vs_L_N}a), we find that starting from sequence lengths of $L=5$ the network works accurately and essentially error-free. It detects the precise moment of deoxidation with an agreement of a few seconds compared to the estimation of the human operator.
We then fix the sequence length to $L=15$ and vary the latent dimension between $N=1$ and $N=70$ (figure~\ref{fig:test_transition} bottom row). For latent vectors of dimension $N=10$ or larger, we find again quasi error-free classification and precise determination of the deoxidation moment.

We conclude that the smallest data size for accurate operation is a latent dimension of $N=10$ and a sequence length of $L=5$. At a video rate of 120 images per full rotation of the substrate holder, this sequence length ($L=5$) corresponds to a rotation angle of 15 degrees being concurrently processed by the classifier network.

\subsubsection{Temporal stability of the classification accuracy}

\begin{figure}[t]
	\centering
	\includegraphics[width=.9\columnwidth]{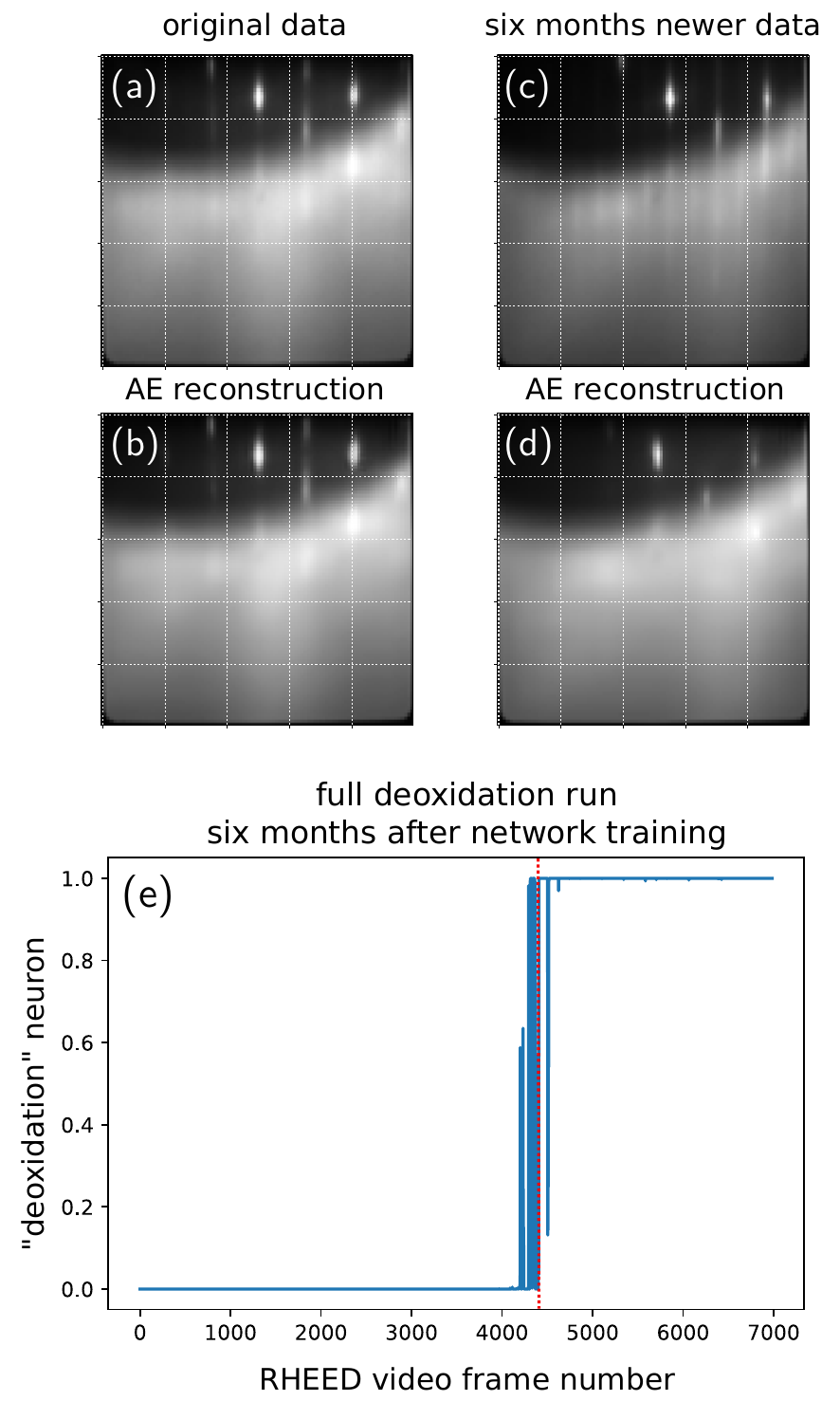}
	\caption{\textbf{Test on six months newer data.} 
		(a) RHEED image of a deoxidized GaAs surface along (1,1,0) incidence, from training data and (b) its reconstruction by the autoencoder with latent vector length $N=50$.
		(c) RHEED image under same incidence angle (along (1,1,0)), but six months later, after having performed more than 200 hours of crystal growth. In particular an increased metallization of the upper third of the RHEED screen is clearly visible, along with a slight displacement of the diffraction pattern due to repeated alignment procedures. (d) its reconstruction by the same autoencoder as used for (b), hence without re-training on new data.
		(e) Evaluation of the RHEED video from a full deoxidation run, recorded 6 months after the training data, without re-training the network model. Latent dimension $N=50$, sequence length $L=15$.
	}
	\label{fig:test_newer_data}
\end{figure}

An MBE is a highly complex apparatus, housing many parts that have a potential impact on the long-term stability of the RHEED precision. To name a few, multiple pumps are required to maintain an ultra-high vacuum, hot source chambers are distributed around the chamber, and mechanical stress is applied to the sample holder. It is not only  constantly rotating but also repeatedly heated up and cooled down by temperature differences of many hundreds degrees Kelvin. Furthermore, frequent (re-)alignment of the RHEED source and camera also have an impact on the reproducibility of the diffraction patterns.
In addition to all these mechanical perturbations, it is very hard to avoid deposition of source material on the RHEED screen, slowly altering the diffraction images during an epitaxy campaign.
In consequence, a constant RHEED image quality can usually not be guaranteed on a long-term.

Deep learning being a data-based technique, a main advantage is generally a high robustness against noise and perturbations, provided that the training data is sufficiently rich \cite{kurumDeepLearningEnabled2019}. 
We therefore expect our approach to deliver reliable classification results over significant time periods.
In order to assess whether our approach can provide accurate deoxidation detection over durations of typical crystal growth campaigns, we recorded a deoxidation RHEED video around 6 months after the acquisition of the training data.

During the 6 months between generation of the training data and this test, around 40 epitaxies, equivalent to roughly 200 hours of growth have been performed on the machine. The most altered component in the system is in fact the RHEED screen, which suffers from metallization over time, implying a considerable deterioration of the RHEED image quality. 
In fact, our test on 6 months newer data was motivated by the observation of a degradation of the RHEED screen, especially in the upper third of the diffraction images (see also Fig.~\ref{fig:setup_scheme}c-d). 

A comparison of ``original'' and ``six months newer'' deoxidized GaAs RHEED images under (1,1,0) incidence is shown in Fig.~\ref{fig:test_newer_data}a and~\ref{fig:test_newer_data}c.
We find that the newer data is not only visually deteriorated as a result of screen metallization (c.f. upper regions of the images). Additionally the newer recordings show diffraction patterns which are shifted with respect to the original training data. The shift is probably due to repeated alignment corrections on the camera and the electron beam positions.

Figures~\ref{fig:test_newer_data}b and~\ref{fig:test_newer_data}d show also the autoencoder reconstructions of the (1,1,0) diffraction patterns respectively from the training set, and six months after. Interestingly, in the reconstruction of the 6 months newer image, the diffraction spots are shifted back to the positions at which they occurred at the time when the original dataset was captured. 
It appears that the convolutional \textit{encoder} stage correctly identifies the information about diffraction in the image. 
The \textit{decoder}, on the other hand, has learned its spatial reconstruction only from the training set, therefore it places the diffraction pattern according to the original RHEED alignment.
In fact, the classifier still works correctly with shifted images. Since classification is done using latent vectors and not with the direct image information, we conclude that the relative diffraction spot / line positions are correctly interpreted by the encoder also from shifted images, and mapped to the same latent variables as with the original training data.

In figure~\ref{fig:test_newer_data}e we show the classification results for a full video from a deoxidation run six months after network training.
While we observe a slightly reduced classification certainty regarding the exact moment of deoxidation, also 6 months after initial training data recording, the non-retrained neural network still performs sufficiently well on the deoxidation detection. 
We want to note that after training the pre-trained network for a few additional epochs on a small set of new images, the fidelity of the network reaches the same confidence as observed in the right-hand panels of figure~\ref{fig:test_transition}.
In conclusion, convolutional neural networks offer a remarkable robustness for data characterization tasks that are susceptible to small perturbations and noise. CNNs are thus particularly interesting for long-term applications in real-world systems.

\subsubsection{Network classification speed}

Deep learning frameworks like tensorflow / keras are highly optimized for batch-processing many samples in parallel. On pre-recorded data, where our models can run parallel data-processing, they are therefore extremely efficient, requiring only a few milliseconds of processing time per sequence. 
In an in-situ operation however, data needs to be processes ``on the fly'', hence image-by-image (or at least one video sequence at a time). For such sequential operation, tensorflow is not optimized. Processing of a single sequence with 15 images by our autoencoder with subsequent classification takes in sum 220\,ms. This is not enough for full real-time analysis of a 24 frames per second video stream as we receive from our RHEED camera. However, the time-scales of MBE crystal growth dynamics are not faster than in the order of tens of seconds, so classification at full video-speed is in fact not required. For the relevant time scales, our method is largely fast enough to practically perform real-time in-situ control. Furthermore the evaluation speed could probably be increased with some further optimization efforts.
Finally, we want to note that our network models are compact enough to be used on conventional CPUs (central processing units). The full processing of a single 15-image sequence takes 220\,ms on our NVIDIA 3070Ti GPU, against 260\,ms on a 10th generation Intel i7 CPU. Hence no additional hardware like graphics processing units is necessary in an MBE control computer.

\section{Conclusions}

In conclusion we presented a deep learning model based on a 2D convolutional neural network to detect the surface oxidation state of GaAs substrates from raw RHEED image sequences, as typically available in molecular beam epitaxy.
Our model consists of a first autoencoder neural network, which learns to compress individual RHEED images to a low-dimensional latent space. 
Sequences of consecutive, compressed RHEED images are then classified for their surface oxidation state through a second convolutional network.
We presented a systematic analysis of classification performance as function of used compression ratio as well as RHEED video sequence length.
We demonstrated that the model accurately identifies the exact surface deoxidation moment and that the performance is robust during at least 6 months of MBE operation without requiring re-training.

While our specific, trained network will of course work only with the MBE setup and RHEED screen used for our training data generation, a generalization to other growth chambers can simply be done by training the same models on according data. Video recording is straightforward, we demonstrated that the approach works well with data from only five deoxidation recordings, and we proved it to function reliably during at least several months.
In consequence, our approach is very appealing thanks to its simplicity and low computational cost. Without requiring additional hardware it can be easily set up in any RHEED equipped MBE.

\begin{acknowledgments}
This work was supported by the Toulouse HPC CALMIP (grant p20010). This study benefited from the support of both the LAAS-CNRS micro and nanotechnologies platform, member of the French RENATECH network, and the EPICENTRE common laboratory between Riber and CNRS.
\end{acknowledgments}


\bibliographystyle{unsrturl}
\bibliography{Ox_Deox_classification.bbl}

\begin{thebibliography}{10}

\bibitem{inoNewTechniquesReflection1977}
Shozo Ino.
\newblock Some {{New Techniques}} in {{Reflection High Energy Electron
  Diffraction}} ({{RHEED}}) {{Application}} to {{Surface Structure Studies}}.
\newblock {\em Japanese Journal of Applied Physics}, 16(6):891, June 1977.
\newblock \href {https://doi.org/10.1143/JJAP.16.891}
  {\path{doi:10.1143/JJAP.16.891}}.

\bibitem{horioNewTypeRHEED1996}
Yoshimi Horio, Yasuyuki Hashimoto, and Ayahiko Ichimiya.
\newblock A new type of {{RHEED}} apparatus equipped with an energy filter.
\newblock {\em Applied Surface Science}, 100--101:292--296, July 1996.
\newblock \href {https://doi.org/10.1016/0169-4332(96)00229-2}
  {\path{doi:10.1016/0169-4332(96)00229-2}}.

\bibitem{braunAppliedRHEEDReflection1999}
W.~Braun.
\newblock {\em Applied {{RHEED}}: {{Reflection}} High-Energy Electron
  Diffraction during Crystal Growth}.
\newblock Springer Tracts in Modern Physics. {Springer Berlin Heidelberg},
  1999.

\bibitem{ichimiyaReflectionHighenergyElectron2004}
A.~Ichimiya, P.I. Cohen, and P.I. Cohen.
\newblock {\em Reflection High-Energy Electron Diffraction}.
\newblock {Cambridge University Press}, 2004.

\bibitem{surfacesSurfaceInterfaceCharacterization1988}
NATO Advanced Study Institute on the~Study of~Surfaces, Interfaces by
  Electron~Optical Techniques, {Valdre}, {U.}, and A.~Howie.
\newblock {\em Surface and Interface Characterization by Electron Optical
  Methods / Edited by {{A}}. {{Howie}} and {{U}}. {{Valdre}}}.
\newblock {Plenum Press : Published in cooperation with the NATO Scientific
  Affairs Division New York}, 1988.

\bibitem{joRealTimeCharacterizationUsing2018}
Janghyun Jo, Youngbin Tchoe, Gyu-Chul Yi, and Miyoung Kim.
\newblock Real-{{Time Characterization Using}} in situ {{RHEED Transmission
  Mode}} and {{TEM}} for {{Investigation}} of the {{Growth Behaviour}} of
  {{Nanomaterials}}.
\newblock {\em Scientific Reports}, 8(1):1694, January 2018.
\newblock \href {https://doi.org/10.1038/s41598-018-19857-2}
  {\path{doi:10.1038/s41598-018-19857-2}}.

\bibitem{thelanderElectricalPropertiesInAs12012}
Claes Thelander, Philippe Caroff, S{\'e}bastien Plissard, and Kimberly~A. Dick.
\newblock Electrical properties of {{InAs1}}-{{xSbx}} and {{InSb}} nanowires
  grown by molecular beam epitaxy.
\newblock {\em Applied Physics Letters}, 100(23):232105, June 2012.
\newblock \href {https://doi.org/10.1063/1.4726037}
  {\path{doi:10.1063/1.4726037}}.

\bibitem{daudinHowGrowCubic1998}
B.~Daudin, G.~Feuillet, J.~H{\"u}bner, Y.~Samson, F.~Widmann, A.~Philippe,
  C.~{Bru-Chevallier}, G.~Guillot, E.~Bustarret, G.~Bentoumi, and
  A.~Deneuville.
\newblock How to grow cubic {{GaN}} with low hexagonal phase content on (001)
  {{SiC}} by molecular beam epitaxy.
\newblock {\em Journal of Applied Physics}, 84(4):2295--2300, August 1998.
\newblock \href {https://doi.org/10.1063/1.368296}
  {\path{doi:10.1063/1.368296}}.

\bibitem{ohtakeStrainRelaxationInAs2020}
Akihiro Ohtake, Takaaki Mano, and Yoshiki Sakuma.
\newblock Strain relaxation in {{InAs}} heteroepitaxy on lattice-mismatched
  substrates.
\newblock {\em Scientific Reports}, 10(1):4606, March 2020.
\newblock \href {https://doi.org/10.1038/s41598-020-61527-9}
  {\path{doi:10.1038/s41598-020-61527-9}}.

\bibitem{bastimanGaAs001Planarization2010}
F.~Bastiman and A.~G. Cullis.
\newblock {{GaAs}}(001) planarization after conventional oxide removal
  utilising self-governed {{InAs QD}} site selection.
\newblock {\em Applied Surface Science}, 256(13):4269--4271, April 2010.
\newblock \href {https://doi.org/10.1016/j.apsusc.2010.02.013}
  {\path{doi:10.1016/j.apsusc.2010.02.013}}.

\bibitem{sachaArtificialIntelligenceNanotechnology2013}
G.~M. Sacha and P.~Varona.
\newblock Artificial intelligence in nanotechnology.
\newblock {\em Nanotechnology}, 24(45):452002, 2013.
\newblock \href {https://doi.org/10.1088/0957-4484/24/45/452002}
  {\path{doi:10.1088/0957-4484/24/45/452002}}.

\bibitem{renModelingApplicationCzochralski2021}
Jun-Chao Ren, Ding Liu, and Yin Wan.
\newblock Modeling and application of {{Czochralski}} silicon single crystal
  growth process using hybrid model of data-driven and mechanism-based
  methodologies.
\newblock {\em Journal of Process Control}, 104:74--85, August 2021.
\newblock \href {https://doi.org/10.1016/j.jprocont.2021.06.002}
  {\path{doi:10.1016/j.jprocont.2021.06.002}}.

\bibitem{schimmelArtificialIntelligenceCrystal2022}
Saskia Schimmel, Wenhao Sun, and Natasha Dropka.
\newblock Artificial {{Intelligence}} for {{Crystal Growth}} and
  {{Characterization}}.
\newblock {\em Crystals}, 12(9):1232, September 2022.
\newblock \href {https://doi.org/10.3390/cryst12091232}
  {\path{doi:10.3390/cryst12091232}}.

\bibitem{choudharyRecentAdvancesApplications2022}
Kamal Choudhary, Brian DeCost, Chi Chen, Anubhav Jain, Francesca Tavazza, Ryan
  Cohn, Cheol~Woo Park, Alok Choudhary, Ankit Agrawal, Simon J.~L. Billinge,
  Elizabeth Holm, Shyue~Ping Ong, and Chris Wolverton.
\newblock Recent advances and applications of deep learning methods in
  materials science.
\newblock {\em npj Computational Materials}, 8(1):1--26, April 2022.
\newblock \href {https://doi.org/10.1038/s41524-022-00734-6}
  {\path{doi:10.1038/s41524-022-00734-6}}.

\bibitem{battieRapidEllipsometricDetermination2022}
A.~Yann Battie, Adri{\`a}~Can{\'o}s Valero, David Horwat, and Aotmane~En
  Naciri.
\newblock Rapid ellipsometric determination and mapping of alloy stoichiometry
  with a neural network.
\newblock {\em Optics Letters}, 47(8):2117--2120, April 2022.
\newblock \href {https://doi.org/10.1364/OL.457147}
  {\path{doi:10.1364/OL.457147}}.

\bibitem{brownModelingMBERHEED1997}
T.~Brown, K.~Lee, G.~Dagnall, R.~Kromann, R.~{Bicknell-Tassius}, A.~Brown,
  J.~Dorsey, and G.~May.
\newblock Modeling {{MBE RHEED}} signals using {{PCA}} and neural networks.
\newblock In {\em Compound {{Semiconductors}} 1997. {{Proceedings}} of the
  {{IEEE Twenty-Fourth International Symposium}} on {{Compound
  Semiconductors}}}, pages 33--36, September 1997.
\newblock \href {https://doi.org/10.1109/ISCS.1998.711537}
  {\path{doi:10.1109/ISCS.1998.711537}}.

\bibitem{vasudevanBigDataReflectionHigh2014}
Rama~K. Vasudevan, Alexander Tselev, Arthur~P. Baddorf, and Sergei~V. Kalinin.
\newblock Big-{{Data Reflection High Energy Electron Diffraction Analysis}} for
  {{Understanding Epitaxial Film Growth Processes}}.
\newblock {\em ACS Nano}, 8(10):10899--10908, October 2014.
\newblock \href {https://doi.org/10.1021/nn504730n}
  {\path{doi:10.1021/nn504730n}}.

\bibitem{provenceMachineLearningAnalysis2020}
Sydney~R. Provence, Suresh Thapa, Rajendra Paudel, Tristan Truttmann, Abhinav
  Prakash, Bharat Jalan, and Ryan~B. Comes.
\newblock Machine {{Learning Analysis}} of {{Perovskite Oxides Grown}} by
  {{Molecular Beam Epitaxy}}.
\newblock {\em Physical Review Materials}, 4(8):083807, August 2020.
\newblock \href {http://arxiv.org/abs/2004.00080} {\path{arXiv:2004.00080}},
  \href {https://doi.org/10.1103/PhysRevMaterials.4.083807}
  {\path{doi:10.1103/PhysRevMaterials.4.083807}}.

\bibitem{kwoenClassificationReflectionHighEnergy2020}
Jinkwan Kwoen and Yasuhiko Arakawa.
\newblock Classification of {{Reflection High-Energy Electron Diffraction
  Pattern Using Machine Learning}}.
\newblock {\em Crystal Growth \& Design}, 20(8):5289--5293, August 2020.
\newblock \href {https://doi.org/10.1021/acs.cgd.0c00506}
  {\path{doi:10.1021/acs.cgd.0c00506}}.

\bibitem{kwoenMulticlassClassificationReflection2022}
Jinkwan Kwoen and Yasuhiko Arakawa.
\newblock Multiclass classification of reflection high-energy electron
  diffraction patterns using deep learning.
\newblock {\em Journal of Crystal Growth}, 593:126780, September 2022.
\newblock \href {https://doi.org/10.1016/j.jcrysgro.2022.126780}
  {\path{doi:10.1016/j.jcrysgro.2022.126780}}.

\bibitem{khaireh-waliehRHEEDDeoxidationDataset2022}
Abdourahman {Khaireh-Walieh}, Alexandre Arnoult, S{\'e}bastien Plissard, and
  Peter~R. Wiecha.
\newblock {{RHEED}} deoxidation dataset, codes and pre-trained models.
\newblock {\em Figshare}, 2022.
\newblock \href {https://doi.org/10.6084/m9.figshare.21695657.v1}
  {\path{doi:10.6084/m9.figshare.21695657.v1}}.

\bibitem{ioffeBatchNormalizationAccelerating2015}
Sergey Ioffe and Christian Szegedy.
\newblock Batch {{Normalization}}: {{Accelerating Deep Network Training}} by
  {{Reducing Internal Covariate Shift}}.
\newblock {\em arXiv:1502.03167 [cs]}, February 2015.
\newblock \href {http://arxiv.org/abs/1502.03167} {\path{arXiv:1502.03167}}.

\bibitem{goodfellowDeepLearning2016}
Ian Goodfellow, Yoshua Bengio, and Aaron Courville.
\newblock {\em Deep {{Learning}}}.
\newblock {MIT Press}, 2016.

\bibitem{springthorpeMeasurementGaAsSurface1987}
A.~J. SpringThorpe, S.~J. Ingrey, B.~Emmerstorfer, P.~Mandeville, and W.~T.
  Moore.
\newblock Measurement of {{GaAs}} surface oxide desorption temperatures.
\newblock {\em Applied Physics Letters}, 50(2):77--79, January 1987.
\newblock \href {https://doi.org/10.1063/1.97824} {\path{doi:10.1063/1.97824}}.

\bibitem{cholletKeras2015}
Fran{\c c}ois Chollet et~al.
\newblock Keras, 2015.

\bibitem{tensorflow2015-whitepaper}
Mar{\'t}{\i}n Abadi, Ashish Agarwal, Paul Barham, Eugene Brevdo, Zhifeng Chen,
  Craig Citro, Greg~S. Corrado, Andy Davis, Jeffrey Dean, Matthieu Devin,
  Sanjay Ghemawat, Ian Goodfellow, Andrew Harp, Geoffrey Irving, Michael Isard,
  Yangqing Jia, Rafal Jozefowicz, Lukasz Kaiser, Manjunath Kudlur, Josh
  Levenberg, Dan Man{\'e}, Rajat Monga, Sherry Moore, Derek Murray, Chris Olah,
  Mike Schuster, Jonathon Shlens, Benoit Steiner, Ilya Sutskever, Kunal Talwar,
  Paul Tucker, Vincent Vanhoucke, Vijay Vasudevan, Fernanda Vi{\'e}gas, Oriol
  Vinyals, Pete Warden, Martin Wattenberg, Martin Wicke, Yuan Yu, and Xiaoqiang
  Zheng.
\newblock {{TensorFlow}}: {{Large-Scale Machine Learning}} on {{Heterogeneous
  Systems}}.
\newblock {\em https://www.tensorflow.org/}, 2015.

\bibitem{bradskiOpenCVLibrary2000}
G.~Bradski.
\newblock The {{OpenCV}} library.
\newblock {\em Dr. Dobb's Journal of Software Tools}, 2000.

\bibitem{waltScikitimageImageProcessing2014}
St{\'e}fan van~der Walt, Johannes~L. Sch{\"o}nberger, Juan {Nunez-Iglesias},
  Fran{\c c}ois Boulogne, Joshua~D. Warner, Neil Yager, Emmanuelle Gouillart,
  and Tony Yu.
\newblock Scikit-image: Image processing in {{Python}}.
\newblock {\em PeerJ}, 2:e453, June 2014.
\newblock \href {https://doi.org/10.7717/peerj.453}
  {\path{doi:10.7717/peerj.453}}.

\bibitem{pedregosaScikitlearnMachineLearning2011}
Fabian Pedregosa, Ga{\"e}l Varoquaux, Alexandre Gramfort, Vincent Michel,
  Bertrand Thirion, Olivier Grisel, Mathieu Blondel, Peter Prettenhofer, Ron
  Weiss, Vincent Dubourg, Jake Vanderplas, Alexandre Passos, David Cournapeau,
  Matthieu Brucher, Matthieu Perrot, and {\'E}douard Duchesnay.
\newblock Scikit-learn: {{Machine Learning}} in {{Python}}.
\newblock {\em Journal of Machine Learning Research}, 12(85):2825--2830, 2011.

\bibitem{collettePythonHDF52013}
Andrew Collette.
\newblock {\em Python and {{HDF5}}}.
\newblock {O'Reilly}, 2013.

\bibitem{fournierEmpiricalComparisonAutoencoders2019}
Quentin Fournier and Daniel Aloise.
\newblock Empirical {{Comparison}} between {{Autoencoders}} and {{Traditional
  Dimensionality Reduction Methods}}.
\newblock In {\em 2019 {{IEEE Second International Conference}} on {{Artificial
  Intelligence}} and {{Knowledge Engineering}} ({{AIKE}})}, pages 211--214,
  June 2019.
\newblock \href {https://doi.org/10.1109/AIKE.2019.00044}
  {\path{doi:10.1109/AIKE.2019.00044}}.

\bibitem{heDeepResidualLearning2015}
Kaiming He, Xiangyu Zhang, Shaoqing Ren, and Jian Sun.
\newblock Deep {{Residual Learning}} for {{Image Recognition}}.
\newblock {\em arXiv:1512.03385 [cs]}, December 2015.
\newblock \href {http://arxiv.org/abs/1512.03385} {\path{arXiv:1512.03385}}.

\bibitem{szegedyInceptionv4InceptionResNetImpact2016}
Christian Szegedy, Sergey Ioffe, Vincent Vanhoucke, and Alex Alemi.
\newblock Inception-v4, {{Inception-ResNet}} and the {{Impact}} of {{Residual
  Connections}} on {{Learning}}.
\newblock In {\em Proceedings of the {{Thirty-First AAAI Conference}} on
  {{Artificial Intelligence}}}, pages 4278--4284, February 2016.
\newblock \href {http://arxiv.org/abs/1602.07261} {\path{arXiv:1602.07261}}.

\bibitem{kurumDeepLearningEnabled2019}
Ulas K{\"u}r{\"u}m, Peter~R. Wiecha, Rebecca French, and Otto~L. Muskens.
\newblock Deep learning enabled real time speckle recognition and hyperspectral
  imaging using a multimode fiber array.
\newblock {\em Optics Express}, 27(15):20965--20979, July 2019.
\newblock \href {https://doi.org/10.1364/OE.27.020965}
  {\path{doi:10.1364/OE.27.020965}}.

\end{thebibliography}

\end{document}